# Simultaneous Interference-Data Transmission for Secret Key Generation in Distributed IoT Sensor Networks

Najme Ebrahimi, *Member, IEEE*, Hun Seok Kim, *IEEE, Member, IEEE,*
David Blaauw, *Fellow, IEEE*

*Abstract:* Internet of Things (IoT) networks for smart sensor nodes in the next generation of smart wireless sensing systems require a distributed security scheme to prevent the passive (eavesdropping) or active (jamming and interference) attacks from untrusted sensor nodes. This paper concerns advancing the security of the IoT system to address their vulnerability to being attacked or compromised by advancement of future supercomputers. In this work a novel embedded architecture has been designed and implemented for distributed IoT network that utilizes a master-slave full duplex communication to exchange the random and continuous modulated phase shift as the secret key to be used in higher-layer encryptions. In the proposed architecture, each IoT node generates a phase modulated random key/data and transmits it to a master node in the presence of an eavesdropper, referred to as Eve. The master node, simultaneously, broadcasts a high power signal using an omni-directional antenna, which is received as interference by Eve. This interference masks the generated key by the IoT node and will result in a higher bit-error rate in the data received by Eve. The two legitimate intended nodes communicate in a full-duplex manner and, consequently, subtract their transmitted signals, as a known reference, from the received signal (self-interference cancellation). We compare our proposed method with a conventional approach to physical layer security based on directional antennas. In particular, we show, using theoretical and measurement results, that our proposed approach provides significantly better security measures, in terms bit error rate (BER) at Eve's location. Also, it is proven that in our novel system, the possible eavesdropping region, defined by the region with BER $< 10^{-1}$, is always smaller than the reliable communication region with BER $< 10^{-3}$. It has been also shown that the security region enhanced significantly (SF<1) in the proposed technique compared to the conventional directional antenna/beam former techniques.

# I. INTRODUCTION

Providing security is a major issue in wireless networks due to their broadcasting nature and the resulting vulnerabilities to eavesdropping attacks. Security is often guaranteed in the higher layers of the network architecture using cryptographic protocols. Such protocols require a secure and random key sequence shared between the authenticated nodes a priori [1], [2]. In contrast, physical layer security methods are keyless and they can be used to securely share random keys to complement the conventional cryptographic techniques [2]. Furthermore, it is well-known that any encryption scheme can be deciphered given a sufficient amount of time and super-computational power. Hence, it is highly desirable to regularly and securely update the shared key in wireless nodes in order to minimize the chances of successful eavesdropping attacks [2], [3].

One previous work for implementing physical layer security is to employ directional antennas that transmit signal using a narrow beam [4]–[6], see Fig. 1(a). To resolve the problem of information leakage in side-lobes a directional modulation technique [5], [6] has been proposed to distort the signal at side-lobes. However, this requires knowledge of the location of the receiver by the transmitter. Furthermore, it has been shown that an eavesdropper, for instance, a small antenna in the main lobe or reflector can detect the signal in the main lobe without degrading the received signal by the intended receiver [7].

In this work, we propose a novel technique to implement physical layer security in the front end. Our approach provides security by broadcasting an intentional interference in a full-duplex scenario that blocks Eve from obtaining the securely generated key, Fig. 1(b). Our protocol does not require any knowledge of the node locations. Also, the proposed architecture does not require directive antennas and only requires omni-directional antennas. We will show that it provides a higher security region ratio compared to previous work such as directional antenna approaches.

In Section II we describe the proposed physical layer security technique. Section III provides the system implementation and the measurement results of this work are presented in Section IV. The paper is concluded in Section V.

## II. MASTER-SLAVE FULL DUPLEX SECURITY TECHNIQUE

### A. Proposed Security Protocol (BER Point of View)

In wireless networks the reliability of communication is often measured in terms of the bit error rate (BER) of the channel. Typically, when BER $< 10^{-3}$, the communication is considered to be reliable. The security level at an unintended receiver Eve is also often measured using the bit error rate. Typically, when BER $> 10^{-1}$ at Eve the communication is considered secure [6]. For instance with BER $> 10^{-1}$ at Eve and assuming a key of length 100, the probability that Eve gets the entire key error-free is $2.7 \times 10^{-5}$.

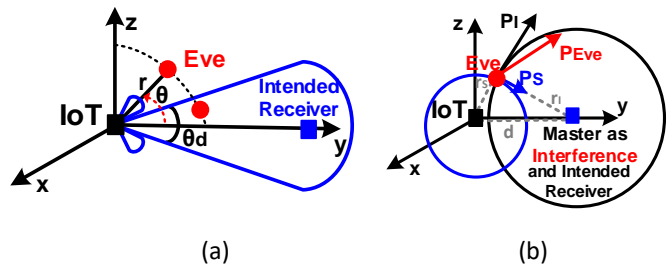

Fig. 1. Conceptual block diagrams of physical layer security approaches. (a) Using directional antenna. (b) The proposed security technique with a master source as interference and an IoT node as secured data transmitter.

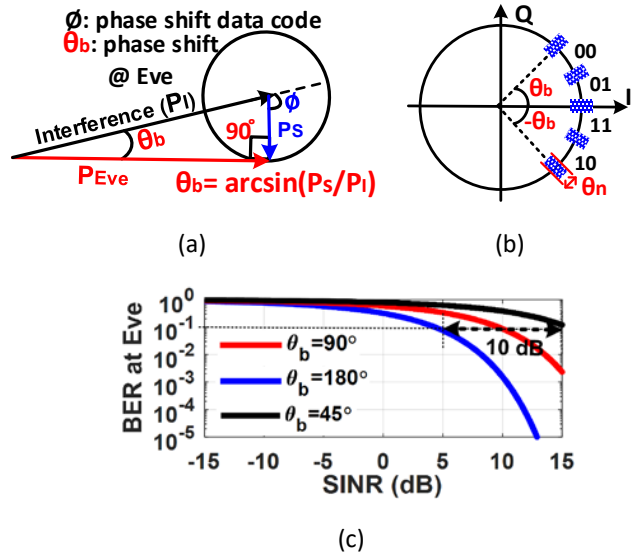

Fig. 2. Proposed technique protocol, (a) interference effect on the maximum phase shift range, $\theta_b$. (b) Conceptual block diagram of key generation protocol, (c) BER versus SINR under various phase shift range, $\theta_b$.

In general, in a phase-based modulation, such as a $M$-ary phase shift keying (M-PSK), bit error rate can be approximated as

$$BER = \frac{1}{k} \cdot 2Q\left(\sqrt{2SINR}\sin\left(\frac{\theta_b}{M}\right)\right) \quad (1)$$

where $M = 2^k$ is the constellation size, $Q(x)$ is the Gaussian complementary error function, $SINR$ is signal to interference- noise ratio and $\theta_b$ is the absolute maximum phase shift range of the modulated data, e.g., 180° for conventional PSK. Then (1) implies that BER increases by reducing the maximum phase shift range $\theta_b$ or by increasing the constellation size.

In our proposed approach, we intentionally reduce the maximum phase shift range at Eve's receiver by broadcasting a high power interference from the master node. Note that Eve receives the spatial summation of the transmitted power by the IoT node and the master node. Then, as illustrated in Fig. 2(a), the maximum phase shift range at Eve occurs when, in the two-dimensional plane, the summation vector is orthogonal to the randomly phase-modulated signal received from the IoT node. Therefore, the maximum phase shift $\theta_b$ at Eve is

$$\theta_{b-EVE} = \arcsin\left(\frac{P_{S@Eve}}{P_{I@Eve}}\right) = \arcsin(SIR_{Eve}) \quad (2)$$

where $P_{S@Eve}$ and $P_{I@Eve}$ are the received power at Eve from the IoT node as the desired secret key and master node as interference, respectively. The ratio of the two is actually the signal to interference ratio at Eve, $SIR_{Eve}$.

Note also that in our approach the phase shift generated by the IoT node is not necessarily constrained by $\pi/M$ as in the conventional $M$-PSK modulations. Instead, the generated phase shift is continuous and random. Then, the maximum phase shift range is split into $M$ regions, each of them corresponding to a key/data in a gray coding format. For example for $M = 4$, as illustrated in Fig. 2(b), the phase shift regions of $(-\theta_b<\varphi<-\theta_b/2)$, $(-\theta_b/2<\varphi<0)$, $(0<\varphi<\theta_b/2)$, and $(\theta_b/2<\varphi<\theta_b)$ corresponds to 00, 01, 11 and 10 key sequences, respectively.

In Fig. 2(c), the BER at Eve is shown for different values of $\theta_b$ at Eve. It can be observed that the SIR or SNR range that satisfy the security condition (BER > $10^{-1}$ at Eve) is improved by 10 dB with $\theta_b = 45°$ comparing to a traditional $M$-PSK with $\theta_b = 180°$.

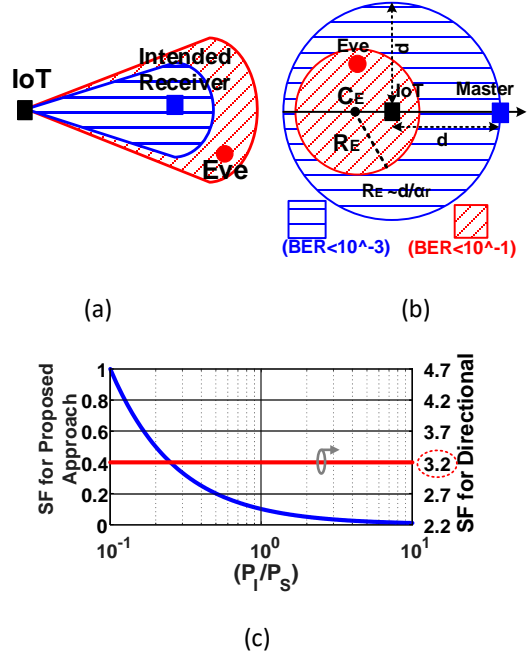

Fig. 3. Reliability and security comparison: (a) directional antenna, (b) proposed approach, (c) security factor comparison.

## B. Comparisons between Security Techniques

Let $P_s$ be the transmitted power by the IoT node, $P_I$ the transmitted power from the master node as undesired interference at Eve (which is proposed in our approach), $P_{N-Eve}$ the overall noise power at Eve, $\gamma_{SE}$ the channel gain between the IoT node and Eve, and $\gamma_{IE}$ the channel gain between the master node and Eve. The channel gain for the directional antenna approach is given by

$$\gamma_{SE} = (\lambda/4\pi)^2 G_s(\theta) G_{r-Eve} \, 1/r^2 \qquad (3)$$

where $\lambda$ is the wavelength, $r$ is the distance of Eve to IoT source, $G_{r-Eve}$ is Eve's antenna gain, assuming an omni-antenna for Eve, and $G_s(\theta)$ is the antenna directive gain. The same equation (3) can be used for $\gamma_{IE}$ and $\gamma_{SE}$ in our approach with $G_s(\theta) = G_s$.

For the directional antenna scheme, there is no interference and the $SINR$ is equal to $SNR$ as:

$$SNR = \frac{P_{S@Eve}}{P_N} = P_S \frac{\gamma_{SE}}{P_N} \qquad (4)$$

In our proposed approach $SINR$ at Eve can be written as:

$$SIR = \frac{P_{S@Eve}}{P_{I@Eve}} = \frac{P_S \gamma_{SE}}{P_I \gamma_{IE}} \qquad (5)$$

Here, the noise power is negligible comparing to the interference power from the master. Using revised (3), (5) can be rewritten as

$$SIR = P_S G_S / P_I G_I \cdot (r_I/r_S)^2 \qquad (6)$$

where $r_S$ and $r_I$ are the distance between Eve and IoT as source and master as interference, respectively, Fig. 1 (b). The $G_S$ and $G_I$ are the normalized antenna gains of the IoT node and the master node, which are omni-directional and can be assumed as identical, respectively.

In order to take into account the condition for reliable communication in our comparison, we define the integrated area regions for both the reliability and security. Let $S_{a-Eve}$ be the eavesdropping region, where BER $< 10^{-1}$ for an Eve node in this region. Similarly, the reliable communication region $S_{a-comm}$ is the region of all locations for the intended receiver with BER $< 10^{-3}$. Then the security factor $SF$ is defined as the ratio of areas of these two regions as

$$SF = \frac{S_{a-Eve}(BER < 10^{-1})}{S_{a-comm}(BER < 10^{-3})} \quad (7)$$

The security factor SF can be used for a fair comparison between different physical layer security techniques. More specifically, given a protocol, a smaller value of $SF$ indicates a higher level of security, in terms of the covered area.

Next, we compute the areas of eavesdropping region, $S_{a\text{-}Eve}$, and communication region, $S_{a-comm}$, for the directional antenna technique and our proposed technique. In the directional antenna approach, the area of region can be expressed as $(\theta_d/2)r^2$, where $r$ is the maximum distance of Eve from the IoT node for a specific probability of error and $\theta_d$ is the directivity angle of IoT antenna. As shown in Fig. 2(c), the constraints (BER $< 10^{-1}$) and (BER $< 10^{-3}$) correspond to $SNR_{min}$ of 10 dB and 15 dB, respectively. Therefore, by (3) and (4), the maximum distance is given by

$$r_{max} \leq \sqrt{\left(\frac{\lambda}{4\pi}\right)^2 \frac{G_s(\theta)G_r}{SNR_{min}} \frac{P_S}{P_N}} \quad (8)$$

Assuming both intended receiver and Eve has same $G_r$ and $P_N$, for communication and the eavesdropping area, the $SF$ ratio of directional antenna technique can be given in terms of SNR as:

$$SF_{dirc} = \frac{SNR_{min}(BER = 10^{-3})}{SNR_{min}(BER = 10^{-1})} \approx \frac{10^{\frac{15dB}{10}}}{10^{\frac{10dB}{10}}} \approx 3.2 \quad (9)$$

Therefore, for the directional antenna scheme, the eavesdropping region is always larger than the reliable communication region, which is shown in Fig. 3(a).

In our proposed technique, $S_{a-Eve}$ is the region where the following condition is satisfied

$$SIR \geq SIR_{min} \overset{(6)}{\leftrightarrow} \frac{r_I}{r_S} \geq \sqrt{\frac{P_I G_I}{P_S G_S} SIR_{min}} \quad (10)$$

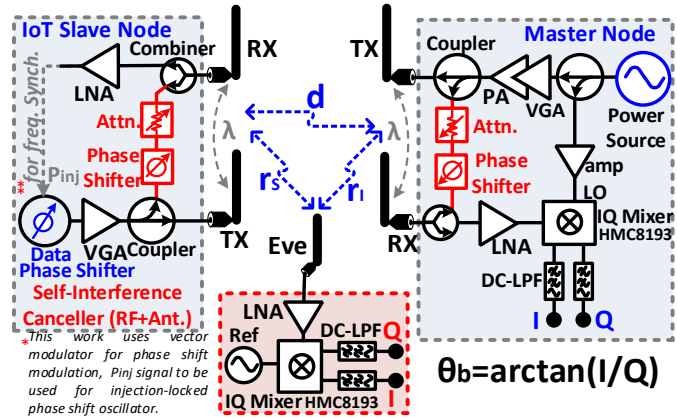

Fig. 4. Block diagram of the proposed system.

where $r_I/r_S$ is the ratio between Eve's distance to the master (interference) and the slave (data source) nodes and $SIR_{min}$ is the SIR at which BER of Eve is $10^{-1}$. Let $\alpha_r = \sqrt{(P_I G_I/P_S G_S)SIR_{min}}$. Then the geometrical representation of $S_{a-Eve}$ described by (10), illustrated in Fig. 3(b), is a circle centered at $C_E$ and with radius $R_E$ considering the data source node as reference of coordinate system

$$R_E = \left|\frac{\alpha_r}{\alpha_r^2 - 1}\right| d \ , C_E = \left|\frac{1}{\alpha_r^2 - 1}\right| d \quad (11)$$

Note that for $\alpha_r \gg 1$ the eavesdropper region is near the IoT node with radius of $d/\alpha_r \ll 1$.

For the reliability condition, i.e., (BER < $10^{-3}$) at the master node, as intended receiver, self-interference cancellation by the master node is the dominating factor. Note that

$$SIR_{@master} = \frac{P_S \gamma_{IS}}{\beta_{si} P_I} \quad (12)$$

where $\beta_{si}$ is the self-interference cancellation at master node and $\gamma_{IS}$ is the channel gain as $\gamma_{IS} = \left(\frac{\lambda}{4\pi d_{max}}\right)^2 G_S G_I$. For a SIR of 15 dB, $P_I/P_S$ of 10 and $\beta_{si}$ of 50 dB, the maximum reliable communication distance, $d_{max}$, of the proposed protocol will be around 1 meter. This can be further improved under enhancement of self-interference rejection. Given the computed radius of the eavesdropping region stated in (11) and the maximum reliable communication distance between the two nodes, $d_{max}$, the security factor for our proposed technique is

$$SF_{prop} = \left|\frac{\alpha_r}{1-\alpha_r^2}\right|^2 \approx \frac{1}{\alpha_r^2} = \frac{1}{SIR_{min}} \frac{P_S G_S}{P_I G_I}. \quad (13)$$

Therefore the security factor of our proposed approach is smaller than 1 under the master slave condition where $(P_I/P_S \gg 1)$. In Fig. 3(c) the security factor of the directional antenna scheme and our proposed approach is shown. It can be observed that our proposed approach is more secure comparing to the directional approach as the eavesdropping region is always smaller than the reliable communication region

## III. SYSTEM IMPLEMENTATION

The block diagram of the proposed system operating at 2.4 GHz is shown in Fig. 4. For the full duplex implementation, two identical omni-directional antennas, $\lambda$ distance apart, are deployed for both Tx and Rx. A vector modulator (HMC631) is used as phase shifter and attenuator for RF self-interference cancellation on both ends, providing more than 50 dB rejection.

In order to have a variable power-interference ratio $(P_I/P_S)$ between the master and the slave, a variable gain amplifier (VGA) (ADL5246) is deployed. The maximum output power by transmitter at the master node varies between -10 dBm and 10 dBm, while it varies at a lower level, between -10 dBm and 0 dBm, at the slave/IoT side. For the Rx path, the master node uses an LNA (PMA-33GLN+) to amplify the received data and to drive the IQ mixer (HMC8193). The reference LO port of the IQ mixer is also driven by a coupled power of master source. Using a DC low pass filter (LPF) (LFCN-160+) the modulated code phase shift can be extracted as $\arctan(I/Q)$. The passive eavesdropper also employs the same IQ mixer with a separate LO reference to extract the phase-modulated key.

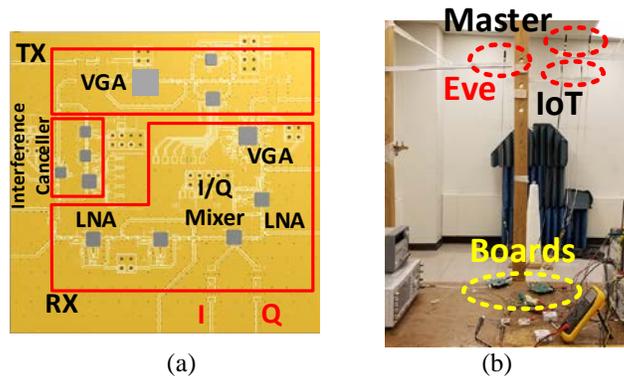

Fig. 5. (a) Board layout of proposed system. (b) Measurement setup.

In order to randomly generate the key a continuous and random phase shift is generated at the IoT node. A vector modulator (HMC631) is used to generate a continuous 360 degree phase shift with variable insertion loss (-51– -11 dB). An injection locked oscillator is an alternative candidate for the proposed system to generate the continuous phase shift, which also enables locking and synchronizing the frequency to the master source. In that case, an LNA can be inserted at the IoT node to amplify the received power from the master source by the injection-locked oscillator for frequency-synchronization [8]. This would also serve as the random phase modulator. This could be a future path of this work together with an IC implementation level.

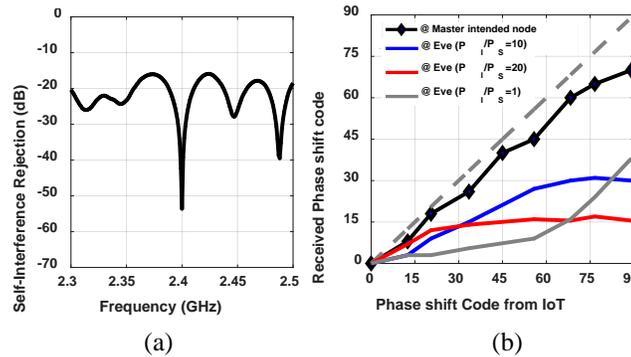

Fig. 6. (a) Measured self-interference rejection at master node. (b) Received phase shift dynamics versus generated phase shift from IoT at master and Eve at $r = d/2$ with different ratio of master-IoT power.

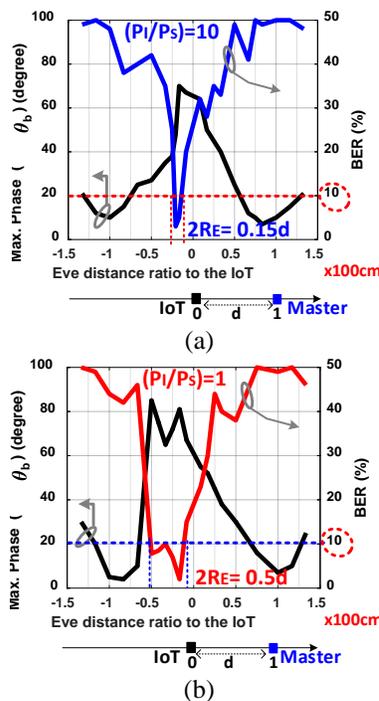

Fig. 7. BER measurement for different Eve location to IoT for $M = 4$. (a) $P_I/P_S = 10$ and (b) $P_I/P_S = 1$.

## IV. Measurement Result

The board is fabricated on FR4 and its layout is shown in Fig. 5(a), which has 8x8cm size. The measurement setup is also shown in Fig. 5(b). The self-interference rejection performance of the master can be tuned from 35 dB to 45 dB rejection, which sets the maximum reliable link distance around 1m, as shown in Fig. 6(a). The maximum received phase shift at the master node while the phase shift at IOT node is continuously varied from 0 to 90° is shown in Fig. 6(b). The maximum received phase shift at Eve at a distance $r = d/2$ from both the master and the IoT node are also shown in Fig. 6(b), for different power ratios, $P_I/P_S$, of 10, 20 and 1. As expected, a larger power ratio of interference significantly reduces the maximum received phase shift.

For measuring the calculated security region, i.e., the radius of the reliable eavesdropping region with BER < 10%, the Eve antenna was located at different distances from the two nodes, including $r_I > d$ and $r_S > d$. The results are shown in Fig. 7(a) and (b) for two different $P_I/P_S$ of 10 and 1, respectively. It can be observed that there is a small distance ratio at which Eve can have a BER smaller than 10% in order to violate the security condition. The distance is equivalent to the radius of the Eve's circle explained in the previous section. This distance is below $0.05d$ for $P_I/P_S$ of 10 and $0.25d$ for $P_I/P_S$ of 1. The measured results of this distance are close to the theoretically computed values presented in (11), which are $R_E = 0.1d$ for $P_I/P_S$ of 10 and $R_E = 0.3d$ for $P_I/P_S$ of 1.

## V. Conclusion

In this work we present a novel technique for physical layer security in the Internet-of-Things (IoT) networks. The proposed architecture uses a master-slave full duplex communication to exchange the modulated random and continuous phase shift as secret key to be used in higher-layer encryptions. As the communication is full duplex the master node can cancel out its self-interference leakage and extract the code transmitted by the IoT. However, this interference will distort the Eve's antenna, preventing it from obtaining an acceptable estimate of the phase shift generated at the IoT.